\documentclass[aps,pra,twocolumn,showpacs,floatfix,preprintnumbers,amsmath,amssymb]{revtex4-2}
\usepackage{amsmath}
\usepackage{amssymb}
\usepackage{graphicx}
\usepackage{epsfig}
\usepackage{dcolumn}
\usepackage{bm}
\usepackage{array}

\begin{document}
\title{Quantum resonances of kicked rotor in the position representation}
\author{Kush Mohan Mittal}
\email{E-mail: kush.mohan@students.iiserpune.ac.in}
\author{M. S. Santhanam}
\email{E-mail: santh@iiserpune.ac.in}
\affiliation{Indian Institute of Science Education and Research,
Dr. Homi Bhabha Road, Pune 411 008, India}

\date{\today}

\begin{abstract}
The study of quantum resonances in the chaotic atom-optics kicked rotor system is of
interest from two different perspectives. In quantum chaos, it marks out the regime
of resonant quantum dynamics in which the atomic cloud displays ballistic mean energy 
growth due to coherent momentum transfer. Secondly, the sharp quantum resonance peaks 
are useful in the context of
measurement of Talbot time, one of the parameter that helps in precise
measurement of fine structure constant. Most of the earlier works rely on fidelity based
approach and have proposed Talbot time measurement through experimental determination of the
momentum space probability density of the periodically kicked atomic cloud. Fidelity approach
has the disadvantage that phase reversed kicks need to be imparted as well which potentially
leads to dephasing.
In contrast to this, in this work, it is theoretically shown that, without manipulating the
kick sequences, the quantum resonances through position space density can be measured more 
accurately and is experimentally feasible as well.
\end{abstract}

\maketitle

\section{Introduction}
Kicked rotor system, a particle that is periodically kicked by an external
sinusoidal field, is a fundamental model of chaotic dynamics in a Hamiltonian system \cite{haake,izr0}.
Atom-optics kicked rotor is an experimentally accessible version of the
quantum kicked rotor model implemented by subjecting cold atomic gases to flashing
optical lattices \cite{moore}. For large kick strengths, classical kicked rotor displays diffusive
mean energy growth while in the corresponding quantum regime it is suppressed by
quantum interference effects leading to dynamical localization \cite{haake,reichl,izr0}. Such distinct dynamical
behaviour in the classical and quantum regime is what makes this system useful to
study the quantum to classical transition and decoherence effects. Recently, atom-optics
kicked rotor has been employed to demonstrate non-exponential coherence decays
and optimal diffusion rates \cite{sumit}. In general, it has been used widely to study unusual
quantum transport scenarios \cite{manai}, quantum entanglement \cite{furuya, arul1, matsui} and in 
quantum metrology for acceleration measurements \cite{rahorne}.

From the point of view of atomic interferometry \cite{horn} and metrology applications, the 
quantum resonances in the atom-optics kicked rotor \cite{izr1, sandro1} 
are of special interest. Such resonances are purely
quantum effects without any classical analogue. In the kicked rotor system, quantum
resonances occur if the scaled Planck's constant is of the form $\hbar=4\pi\frac{l}{s}$, where $l$ and $s$ are integers. For $s=1$, time interval between successive kicks is called the Talbot time and the energy
of the kicked rotor system grows quadratically due to coherent momentum transfer to
the atomic cloud. However, if $s\neq1$, then an initial state recurs after $s$ kicks,
in which the phase added by successive kicks tends to cancel each other. This is called 
the anti-resonances \cite{dana}. Measurement of Talbot
time through resonance effects is important for determining precise values of fine 
structure constant and hence a crucial ingredient for quantum metrology \cite{qmetro}.
The quantum resonances are at the heart of atom interferometers and
were experimentally realised using atom-optics kicked rotors \cite{tony1, ryu, lepers2}.
One proposal for atom-optics kicked rotor based inteferometry relies on manipulating
kick sequences in such a way that the $N$ pulses imparted to the atomic cloud are
followed by $N$ pulses whose phases differ by $\pi$ with respect to the former 
sequence \cite{daszuta}. This is shown to be capable of measuring Talbot time 
and the local gravitational field \cite{daszuta}.
It is also known that quantum resonances are reasonably robust to small amounts of 
phase noise \cite{white} and amplitude noise \cite{sadgrove,brouard} in the kick sequences, thus 
making these systems attractive for interferometric studies and measurement of Talbot time.

Atom optics kicked rotor system can be used to measure Talbot time and it mainly relies on
momentum space measurements. The first order effects of perturbation about the Talbot time 
shows up as a change in the phase of the coefficients in momentum basis. Hence, this effect
cannot be detected directly by measuring the momentum distribution
of atoms \cite{mcdowall, daszuta}.
To measure these effects, a fidelity measurement has been proposed involving the
application of phase reversed kicks \cite{mcdowall}.
In this scheme, $N$ periodic kicked rotor pulse sequences of strength $k$ are followed by
the last pulse which will have opposite phase and strength $N k$. This is shown to display
a sharp resonance peak whose width is proportional to $N^{-3}$ \cite{mcdowall} in contrast
to mean energy resonant profiles which scales as $N^{-2}$.
This scenario was experimentally observed in a fidelity measurement performed on
Bose-Einstein condensate in pulsed external field \cite{taluk}. Although this was able to 
capture the first order changes in the phase, experimentally the kick reversal process 
led to pronounced dephasing. Hence this scheme was not feasible for
large kick numbers. Thus, dephasing became an impediment to achieving even sharper resonance
peaks. 

\begin{figure}[t]
\centering
\includegraphics[width=3.5in]{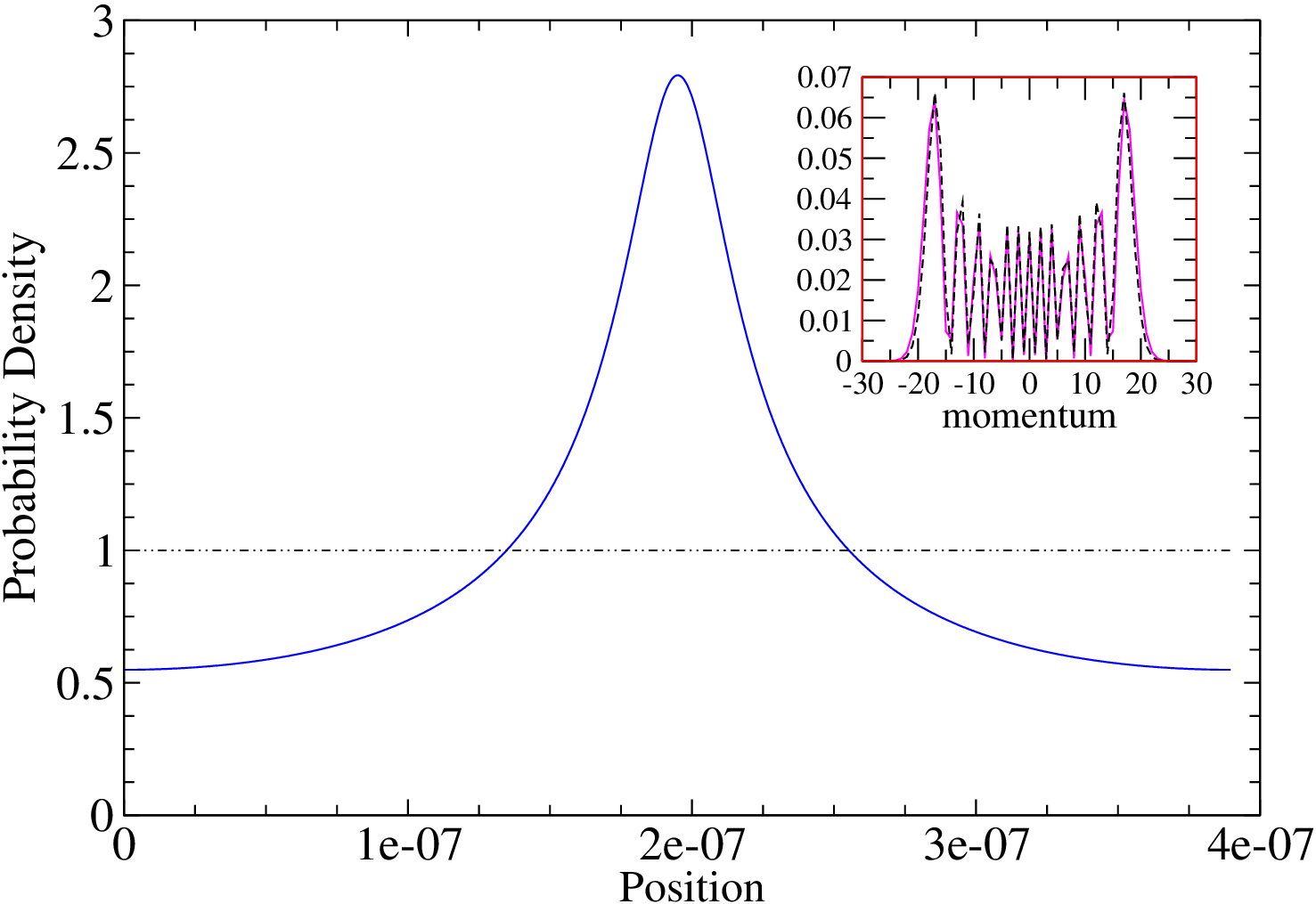}
\caption{Position space density after $N=40$ kicks have been imparted to the initial state
evolving under the effect of kicked rotor Hamiltonian.
The dotted line corresponds to $\varepsilon=0$ and solid line to $\varepsilon=10^{-8}$.
The inset shows the momentum space density for the same values of $\varepsilon$ as
the main figure. Note the pronounced difference between the cases of these two values
of $\varepsilon$ observed in the position representation, but not in the case 
of momentum representation.}
\label{pmdiff}
\end{figure}

In order to overcome this deficiency, in this paper, we propose position space 
measurement and this does not require manipulating the standard pulse sequence imparted
to the kicked rotor system. This is motivated by the fact that if the kick period differs from 
the Talbot time by $\varepsilon \ll 1$, then to first order, the resulting effect
shows up directly in the spatial distribution of atoms as a narrower initial peak 
about the Talbot time without the requirement of the kick reversal process. On the other
hand, in the momentum space, the effect is not distinct from the case with $\varepsilon=0$.
Hence, quantum resonance is better analysed in the position representation.
This is illustrated in Fig. \ref{pmdiff}. In this, the position space density is displayed
for fixed values of $\varepsilon=0$ and $10^{-8}$ after $N=40$ kicks are imparted.
Clearly, in comparison with the case of $\varepsilon=0$, for small deviation from
Talbot time, position density shows pronounced difference. In the momentum representation
(shown in the inset of Fig. \ref{pmdiff}), the probability densities for $\varepsilon=0$ and 
$10^{-8}$ after $N=40$ kicks do not show any perceptible difference. The analysis
presented in this paper is motivated by this observable effect in position space as well
as the fact that the experiments using optical mask techniques can directly probe position 
space density of cold atomic cloud \cite{turlapov}.
Further, we also extend the results in Ref. \cite{lepers} to derive analytical 
result for change in position density around the Talbot time. We compare the analytical
results with the simulations.

\section{Quantum resonances in the $\delta$-kicked rotor}
Kicked rotor system is experimentally realized using ultra-cold atoms in optical lattices.
In this, the periodic kicks are imparted using two far-detuned counter propagating laser beams. 
The pulse are considered to be in the short pulse or Raman-Nath limit, {\it i.e.}, the evolution
of the atomic cloud during the pulse duration is negligible. In the ideal limit of delta 
kicks the system is described by the dimensionless Hamiltonian \cite{reichl} given by
\begin{equation}
    \widehat{H}=\frac{\widehat{P}^2}{2M}+K \cos\widehat{X} ~ \sum_{n=0}^{N-1}\delta(t-n).
\end{equation}
where we have assumed that a total of $N$ kicks are imparted to the atomic cloud.
In this, $K$ is the kicking strength, $k_L$ is the wave number of the standing wave,
$T$ is periodicity of the kicks. Both $\hat{X}$
and $\hat{P}$ are the scaled canonical variables.
The corresponding evolution operator is
\begin{equation}\label{Eq.(2)}
    \widehat{U} = \exp\left(-i\frac{K}{\hbar_s} \cos X \right) \exp \left(-i\frac{P^2}{2 \hbar_s} \right),
\end{equation}
in which $\hbar_s=4\hbar k_L^2T/M$ represents the scaled Planck's constant. 
The evolution operator splits into the kick and the free evolution part due to the $\delta$-kicks. 
As the kicking potential is spatially periodic, the eigenstates have a Bloch wave
structure. Hence, an initial state $|P_o\rangle$ corresponding to definite momentum,
under the action of the evolution operator, gets mapped to states of the form $|P_o+m \hbar_s \rangle$, 
with $m$ being an integer.  In the analysis presented below, we shall be considering the case 
where the particles are initially in the zero momentum state, i.e, $|P_o=0\rangle$. It must be noted
that if the profile of the initial state has a width due to finite temperature effects, then
the temporal evolution of the initial state strongly deviates from that corresponding
to $|P_o=0\rangle$ for longer times. This time-scale is inversely proportional to the
width of the initial momentum distribution \cite{saunders}. In experiments, at longer times 
(more kicks) the dephasing effects become significant as well. Hence, the analysis 
presented here is valid until dephasing becomes pronounced even if initial states are not ideal. Further, the results of numerical simulations shown in this
were performed for an effective kick strength of $K/\hbar_s = 0.485$ and $\lambda_L$
corresponding to a wavelength of 780 nm.

Considering this, we can write the state of the particle at any arbitrary time in the form,
\begin{equation}\label{Eq.(3)}
    \Psi(X)=\frac{1}{\sqrt{2\pi}}\sum_{m=-\infty}^{\infty}\psi(m)~e^{imX}.
\end{equation}
Quantum resonance is characterized by the value of $\hbar_s=4\pi l$, $l > 0 $ is an integer. 
Then, the free evolution part of $\hat{U}$ given in Eq.\eqref{Eq.(2)} becomes $\exp(-i 2 \pi m^2 l)$,
effectively an identity operator. For this choice of $\hbar_s$, the kick period $T$ is referred
to as the Talbot time $T_b$ corresponding to the physical picture that the kicks accumulate
phases in a coherent manner \cite{lepers}. In the analysis that follows, we compute the
change in $\Psi(X)$ due to small perturbation in the Talbot time $T_B$. Unlike the earlier
studies in momentum representation, we study the resonances in position representation. 

\section{Perturbation about the Talbot time}

\begin{figure}[t]
\centering
\includegraphics[width=3.5in]{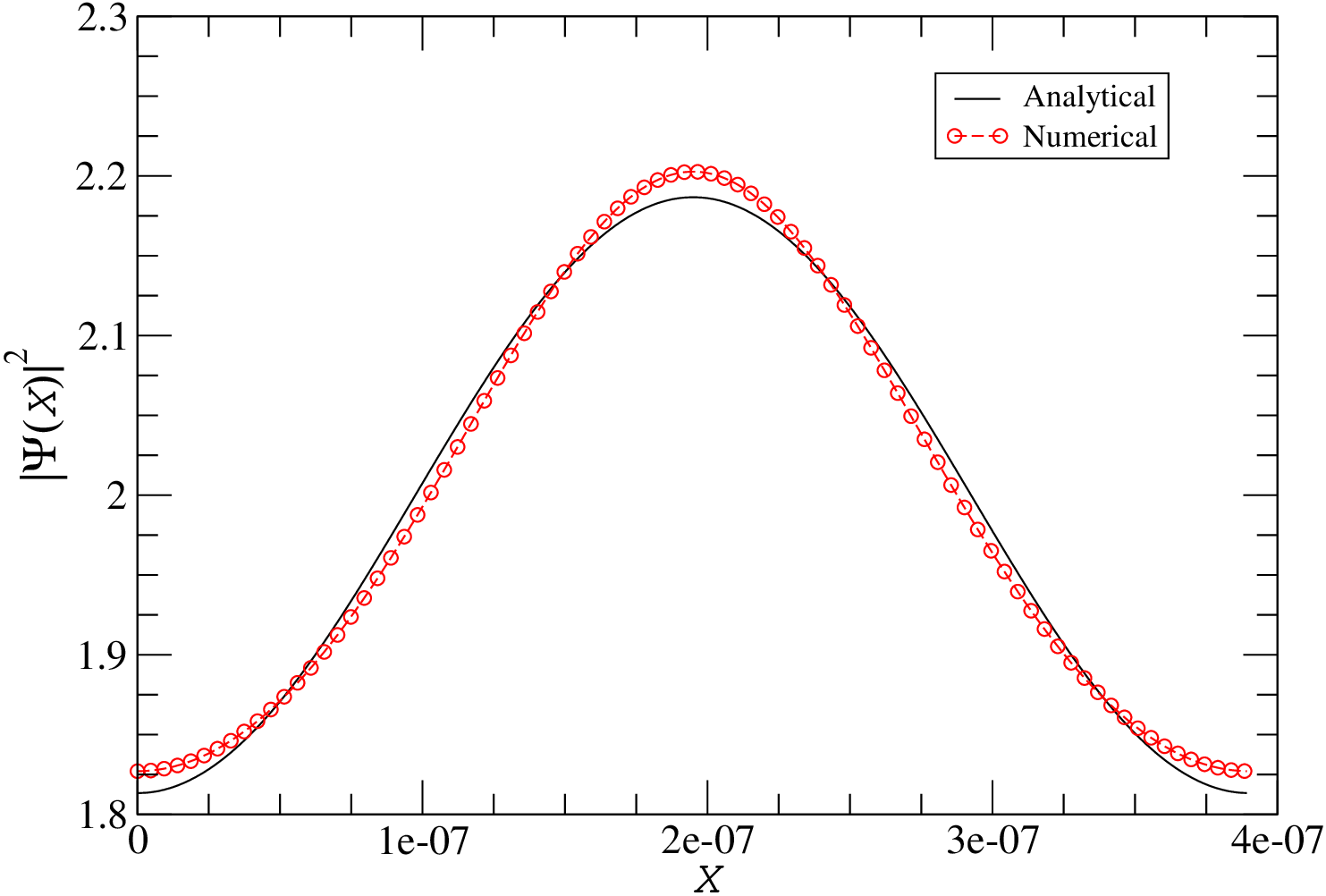}
\caption{Position space density after $N=5$ kicks have been imparted to the initial state
evolving under the effect of kicked rotor Hamiltonian. The deviation from Talbot time
is $\varepsilon=10^{-7}$. The solid curve is the 
analytical result in Eq. \ref{Eq.(16)} and symbols are the numerically computed result.}
\label{probdenX}
\end{figure}

As pointed out earlier, we assume that only $N$ kicks are to be imparted. In the experimental
context, depending on the parameters of the set-up, typically the number of kicks do not exceed 
a few hundred in units of kick period.
Let $\Psi(X,t-1)$ represent the state at time $t-1$ and corresponds to the time when the 
$(N-1)^{th}$ kick is just applied. Following this, the free evolution operator is applied to
it for time $t=T_B + \varepsilon$ where $\varepsilon \ll 1$ is a small perturbation about 
the Talbot time. This leads to
\begin{equation}
    \Psi(X,t^-) = \exp\Big(-i\frac{P^2}{2\kappa}\Big)\Psi(X,t-1),
\label{psitminus}
\end{equation}
in which $t^-$ denotes the state just before the $N^{th}$ kick is applied. 
Now, as discussed earlier, $P=m\hbar_s$ and it is evident that for small perturbations
about the Talbot time the scaled Planck's constant becomes $\hbar_s=4\pi l (1+  \frac{\varepsilon}{T_B})$.
Substituting these values for $P$ and $\hbar_s$ in Eq.(\ref{psitminus}), we get
\begin{equation}
    \Psi(X,t^-)=\exp\left(-i\frac{m^2}{2} 4\pi l (1+  \frac{\varepsilon}{T_B})\right)\Psi(X,t-1).
\end{equation}
The exponential term can be split into two parts; one corresponding to the ideal Talbot time 
condition and other corresponding to the perturbation. For $\varepsilon \ll 1$, the exponential
is expanded as a Taylor series and terms of order $\varepsilon^2$ and higher are discarded.  
We consider the primary resonance by taking $l=1$, and this leads to
\begin{align}
    \Psi(X,t^-)&=e^{-i2\pi m^2}\Psi(X,t-1) \exp\Big(-i2m^2\pi\frac{\varepsilon}{T_B}\Big) \nonumber \\
               &=\Psi(X,t-1)-i2m^2\pi\frac{\varepsilon}{T_B}\Psi(X,t-1)
\end{align}
Using Eq.\eqref{Eq.(3)}, $\Psi(X,t^-)$ can be rewritten as
\begin{align}
    \Psi(X,t^-) = & \Psi(X,t-1) ~ - \label{9}  \\
                  & \frac{1}{\sqrt{2\pi}}\sum_{m=-\infty}^{\infty}\psi(m,t-1)e^{imX}
                                  \Big(i2m^2\pi\frac{\varepsilon}{T_B}\Big) \nonumber
\end{align}
To analyze the first order effect of the perturbation, the position space density 
is obtained as
\begin{align}
    |\Psi(X,t^-)|^2 & =  |\Psi(X,t-1)|^2 - \frac{2\pi i \varepsilon}{\sqrt{2\pi}T_B}  \nonumber \\
                    & \sum_{m=-\infty}^{\infty} m^2 \left[ \Psi^*(X,t-1)\psi(m,t-1)e^{imX} ~ +\right.  \nonumber \\
                    & \left. \Psi(X,t-1)\psi^*(m,t-1)e^{-imX} \right], 
\label{psi2}
\end{align}
in which only the terms first order in $\varepsilon$ are retained. The first term on the right hand side
corresponds to the probability density at time $t-1$ which already includes the effect of
small perturbation about Talbot time. The rest of the terms are first order corrections 
at $N$-th kick. Using Eq.(\ref{Eq.(3)}) again, the correction terms appearing in Eq.(\ref{psi2}) can
be written as
\begin{equation}
\begin{split}
    \sum_{n=-\infty}^{\infty}\sum_{m=-\infty}^{\infty} \frac{i m^2 \varepsilon}{T_B}
\Big( \psi^*(n,t-1)\psi(m,t-1)e^{i(m-n)X}  + \\
\psi(n,t-1)\psi^*(m,t-1)e^{i(n-m)X} \Big).
\end{split}
\end{equation}
It is evident that the terms for which $n=m$ cancel each other and the summation
is left with terms with $n\neq m$. Thus, the corrections can equivalently be written as,
\begin{equation}
\label{Eq.10}
\begin{split}
    \sum_{m=-\infty}^{\infty}\sum_{n>m}2 ~Re\Big[\frac{1}{{2\pi}}\psi^*(n,t-1)\psi(m,t-1)e^{i(m-n)X} \\
             \Big(i(n^2-m^2)\frac{2\pi\varepsilon}{T_B}\Big)\Big]
\end{split}
\end{equation}
where $Re(.)$ represents the real part. Since we are working upto first order in $\epsilon$, the amplitude $\psi$ in Eq.(\ref{Eq.10}) corresponds
to that when the Talbot condition is met. Hence they can be written in terms of $n$-th order
Bessel function $J_n(.)$ as $\psi(n,t-1)=(-i)^nJ_n((N-1)\phi_d)$ \cite{cuyt} where $\phi_d=K/\hbar_s$ at time $t-1$ when the $(N-1)^{th}$ kick has been applied. This leads to the expression
for the correction term $C_N(\varepsilon)$ :
\begin{align}
C_N(\varepsilon) & = \frac{1}{\pi} \sum_{m=-\infty}^{\infty}\sum_{n>m}  
Re \Big[ e^{i(m-n)X} (n^2-m^2)\frac{2\pi i \varepsilon}{T_B} \nonumber \\
                 & i^n J_n((N-1)\phi_d) (-i)^m J_m((N-1)\phi_d) \Big]
\end{align}
In this, $J_m(.)$ is the Bessel function of order $m$ with real argument.
Thus, the probability density under the first order approximation can be 
finally written as,
\begin{equation}
 |\Psi(X,t^-)|^2=|\Psi(X,t-1)|^2 + C_N(\varepsilon).
\label{Eq.(16)}
\end{equation}
Physically, the effect of kick is to give a phase factor of 
$\exp(-i\frac{K}{\hbar_s}\cos X)$. This does not affect the probability density 
in position basis, but leads to occupancy of higher momenta states which in 
turn affects the perturbation term.
It is clear that Eq.\eqref{Eq.(16)} forms a recursive equation connecting $\Psi$
at successive kicks and complete analytical though cumbersome approximation can 
be found recursively.

\section{Position space analysis}
In Fig. \ref{probdenX}, perturbation based analytical result obtained in Eq.(\ref{Eq.(16)}) is compared
with the numerical simulations for $\varepsilon=10^{-7}$ and $N=5$ kicks. The analytical 
result is calculated recursively starting from momentum eigenstate $p_0=0$ at $t=0$. The
kicked rotor Hamiltonian is evolved with kick strength $\phi_d=K/\hbar_s=0.485$.
There is a reasonably good agreement between the analytical and the numerical position
space distribution. In particular, the quantity of interest, the standard deviation of the distribution, is
well captured by the analytical result.

\begin{figure}[t]\centering
\includegraphics[width=8.5cm]{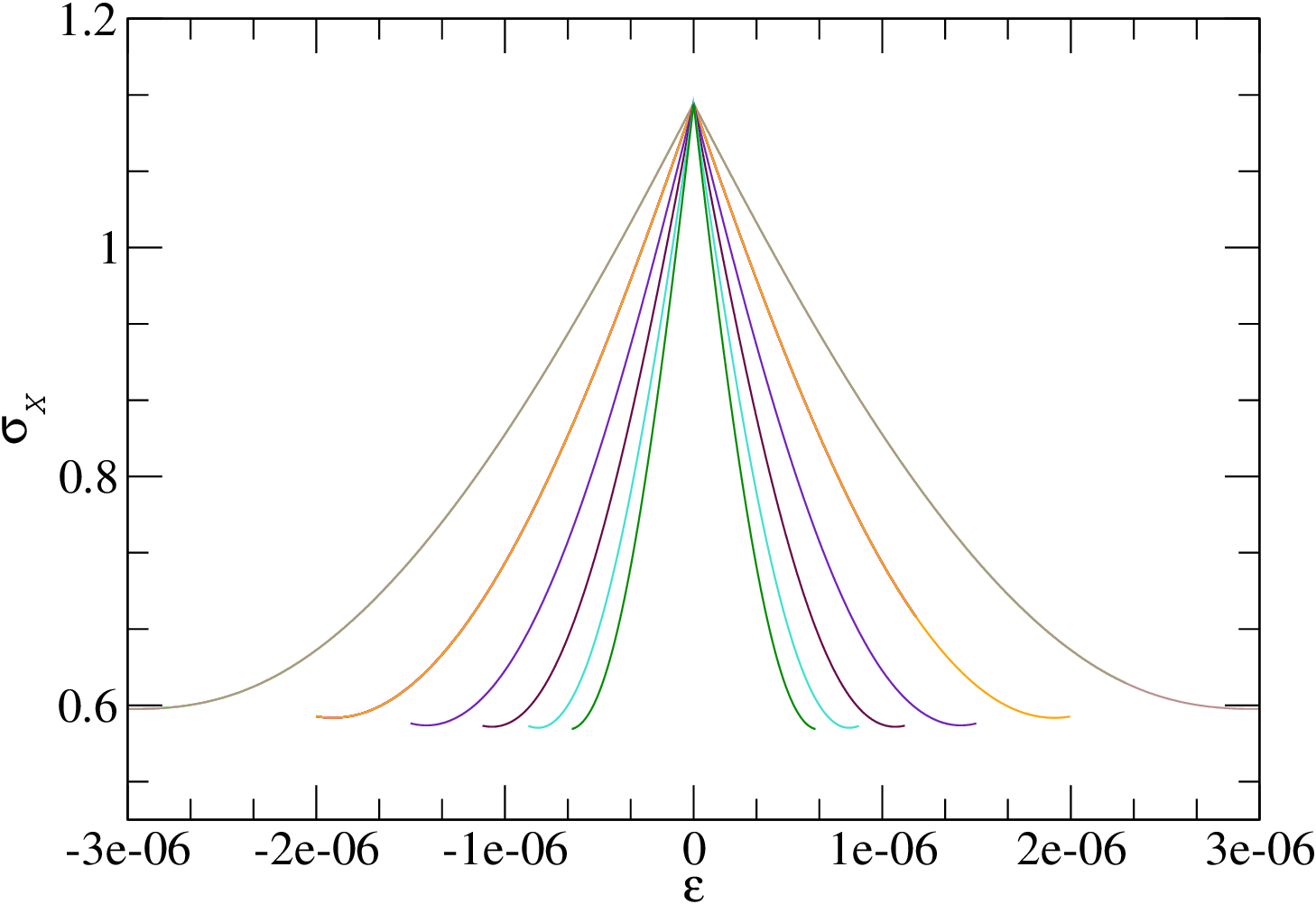}
\caption{The numerically computed standard deviation $\sigma_X$ of the probability density function as a 
function of perturbation about Talbot time $\varepsilon$ for a fixed kick number. 
It is displayed for the kick numbers $N=5$ (outer most curve) to $N=10$ (inner most curve).
The value of $\sigma_X$ decreases with increase in the kick number.}
\label{fwhm}
\end{figure}

For the present purposes, the quantity of interest would be the sharpness of the 
resonance in position space density $|\Psi(X)|^2$. This is conveniently measured using the 
standard deviation denoted by $\sigma_X$. It is anticipated that as the kick period
deviates from Talbot time, as quantified by $\varepsilon$, the position space density
will evolve from an uniform profile at $N=0$ (shown in Fig. \ref{pmdiff})
to a narrow profile as $N \gg 1$. The width $\sigma_X$ of this profile will decay with 
increasing kick number. As is evident from the numerical results in Fig. \ref{fwhm}, as the number of kicks increase, 
the width does indeed decrease. It takes a power-law form $\sigma_X \propto N^{-\gamma}$, and
the exponent $\gamma$ is estimated by regression to be 2.10.

It must be noted that the width in the case of fidelity based analysis scales with $N$
whose exponent is $\gamma \approx -3.0$ \cite{mcdowall}. Thus, approximately, the width
of the distribution for position space analysis scales as $1/N^2$ whereas for fidelity
analysis it scales as $1/N^3$. Notice also that in general the width of position space
based method starts from far lower width in comparison to the fidelity method.
It turns out that till $N=16$ kicks, the position space distribution has lesser $\sigma$
than that of fidelity approach and hence can potentially lead to better Talbot time measurement. 

\begin{figure}\centering
\includegraphics*[width=8.5cm]{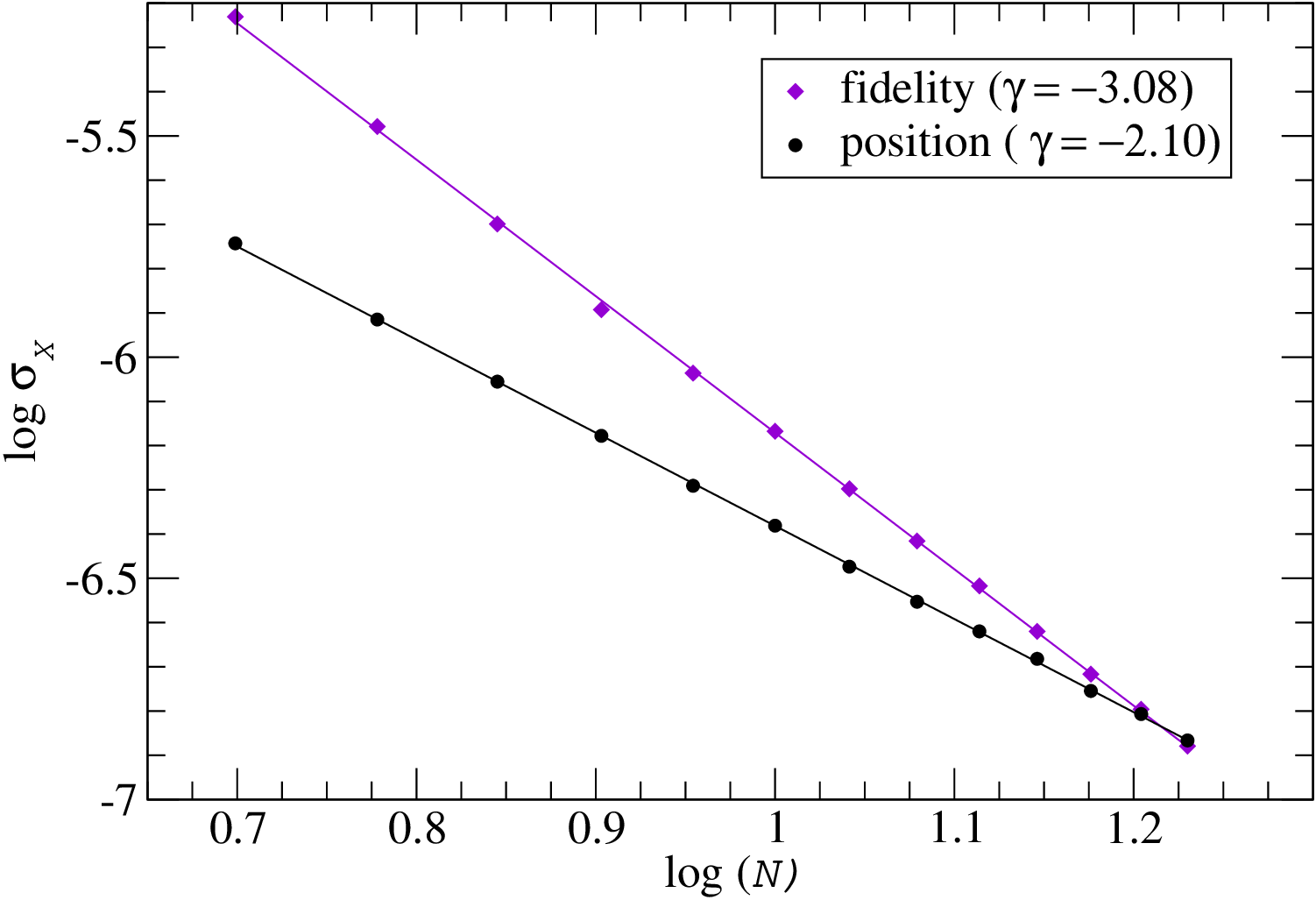}
\caption{The width $\sigma$ of distribution about the Talbot time as a function of
kick number for fidelity analysis (reported in \cite{mcdowall}) and the position 
space analysis studied in this paper. The symbols are the simulation results and the
solid lines represent best-fit lines. The slope for fidelity analysis is estimated
to be $\gamma = -3.08$ while for the position space analysis it is $\gamma=2.10$.}
\label{fwhm_comp}
\end{figure}

However, one significant problem with the fidelity technique is the requirement of
kick reversal process, which can potentially lead to dephasing. This implies that, for
a fair comparison of both these approaches, if position space analysis presented here
applied $M$ kicks, then fidelity technique will require $2M$ kicks, {\it i.e.}, $M$ normal kicks
plus $M$ phase reversed kicks, to be applied. If this is taken into account, then for
identical {\it total} number of kicks applied, it is 
clearly seen that position space density based analysis of quantum resonance
far outperforms the fidelity based approach.

\section{Conclusions}

The quantum resonances in the position space representation
of the atom optics kicked rotor system is analysed in this paper.
In the kicked rotor system, quantum resonances can be observed when the scaled Planck's constant is given by, $\hbar=4\pi l$, where $l$ 
is an integer. Thus, this analysis provides theoretical results
for the measurement of Talbot time in position space representation.
Generally, the atom-optics based kicked rotor experiments perform
measurements in momentum space. However, the central result of this
paper is that the quantitative signatures of quantum resonances, and 
hence the Talbot time, are better inferred from position space, 
rather than in momentum space, representation.

For this purpose, we have analytically obtained the first order changes 
in the position space density of the evolving atomic cloud about Talbot time. 
The basic idea in this work is to capture the changes which occur in the
phase in momentum space directly by measuring the probability distribution in 
position space. This can be further extended to various other setup and problems 
as an efficient alternative to fidelity based treatment in order 
to capture the phase information. The latter requires kick sequence manipulation
which often lead to dephasing effects of the atomic cloud within a few kicks.
It is shown that the position space analysis can be more accurate in measuring 
the Talbot time. Further, it is much more experimentally feasible than the fidelity 
treatment quantum resonances based on momentum space analysis. In this position
representation approach, the width of the quantum resonances scale approximately
as $N^{-2}$, where $N$ is the number of kicks. Significantly, this does not 
require manipulating the kick sequence (as is normally done for fidelity 
based approaches). As experimental techniques for directly probing position
space density through optical mask techniques are available \cite{turlapov}, 
this work might lead to direct Talbot time measurements.



\end{document}